\input{aipcheck}

\documentclass[final]{aipproc}

\layoutstyle{8x11single}

\begin{document}

\title{Symmetry Tests in Photo-Pion Production }


\author{A.M.Bernstein}{
  address={  Physics Dept. and Laboratory for Nuclear Science\\
         Massachusetts Institute of Technology\\
         Cambridge MA. 02139, USA \\  
       email= {bernstein@mit.edu} }
        }

\begin{abstract}
Small angle electron scattering with intense electron beams opens up the possibility of performing almost real photon induced reactions with thin, polarized hydrogen and few body targets, allowing for the detection of low energy charged particles.This promises to be much more effective than conventional photon tagging techniques. For photo-pion reactions some fundamental new possibilities include: tests of charge symmetry in the N-N system by measurement of the neutron-neutron scattering length $a_{nn}$ in the $\gamma D \rightarrow \pi^{+} nn$ reaction; tests of  isospin breaking due to the mass difference of the up and down quarks; measurements with polarized targets are sensitive to $\pi$N phase shifts and will test the validity of the Fermi-Watson (final state interaction) theorem. All of these experiments will test the accuracy and energy region of validity of chiral effective theories.
\end{abstract}
\maketitle





The development of low energy, high intensity electron accelerators raises the possibility of conducting fundamental experiments in photo-pion production from the proton and neutron (in few body targets)  by detection of small angle electro-pion production at very low $Q^{2}$ values. This  builds on  the demonstrated value of high current electron scattering experiments with thin, polarized  targets  by the BLAST experiment at Bates\cite{BLAST}. It was a planned future project at Bates\cite{POLITE} which was not carried out due to the termination of funding. The method will first be presented followed by a few important possibilities.

Forward angle electron scattering (virtual photon tagging) is potentially significantly more effective than conventional photon tagging and  will open up significant experimental possibilities. The reason is that all of the detected electrons will have interacted in the target and have lost energy. This allows for the use of higher currents, thin targets and the detection of low energy changed particles. This can also allow the utilization of atomic beam polarized targets which are isotopically pure but thin. In a conventional tagger the photons are produced in an upstream radiator which is limited by the maximum allowable rate in the tagger detectors (typically $\leq$ 1 MHz). Most of the tagged  photons do not interact in the target. This loss is compensated for by the use of thick targets which do not allow for the detection of low energy charged particles (For the latest photo-pion production experimental results and references to tagged photons see\cite{Hornidge}). In addition, to obtain polarized targets one must use complex materials such as butanol (H10,C4,O) which produce large backgrounds due to the heavier elements.

\begin{figure}
  \includegraphics[height=.25\textheight, width = 0.55\textwidth]{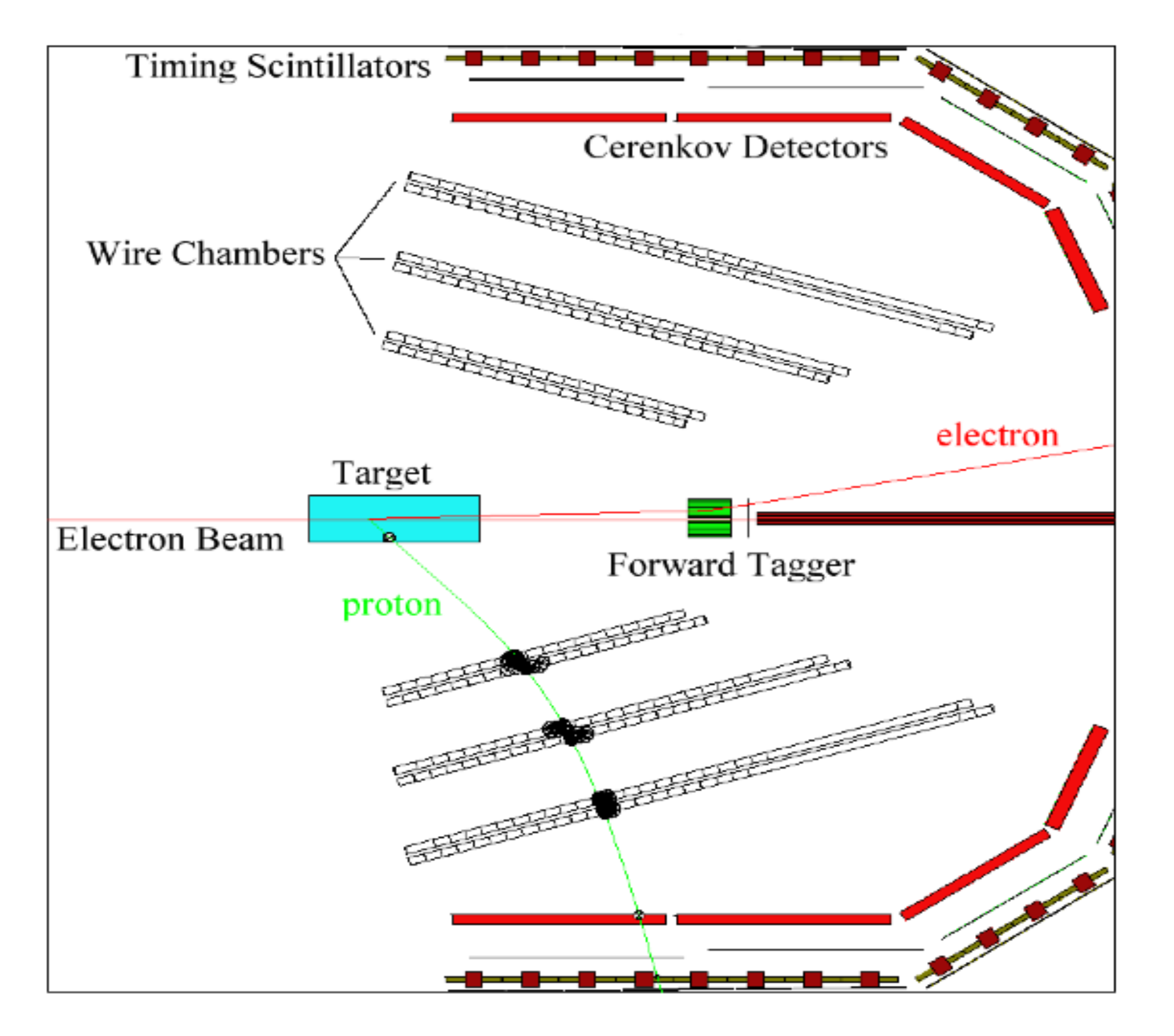}
  \includegraphics[height=.25\textheight]{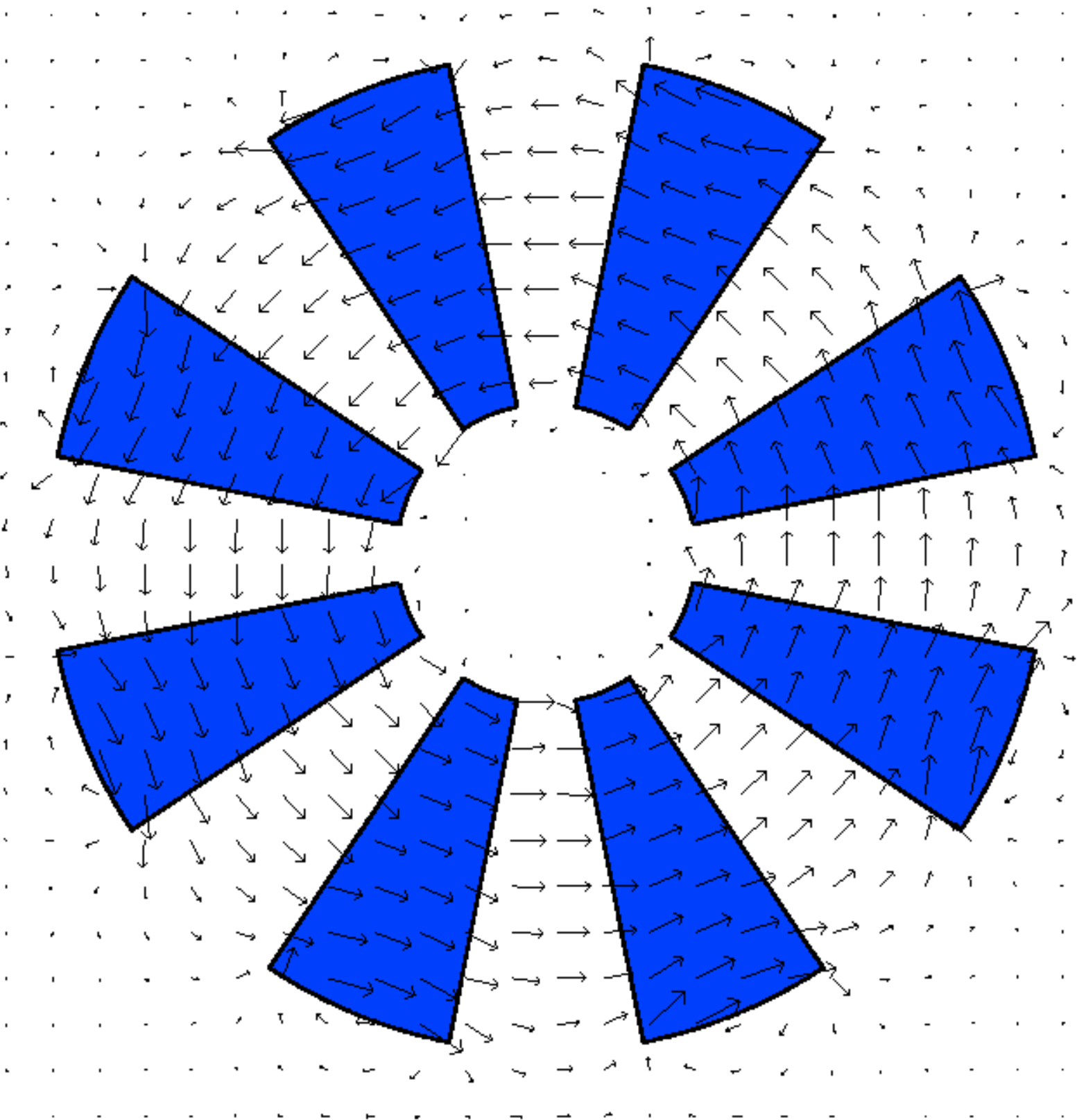}
  \caption{Schematic internal target setup.Left panel: Target and detectors in a magnetic field. Right panel: forward tagger magnet. See text for discussion  }
  \label{fig:setup}
\end{figure}

\begin{figure}
  \includegraphics[height=.3\textheight,width = 0.45\textwidth]
   {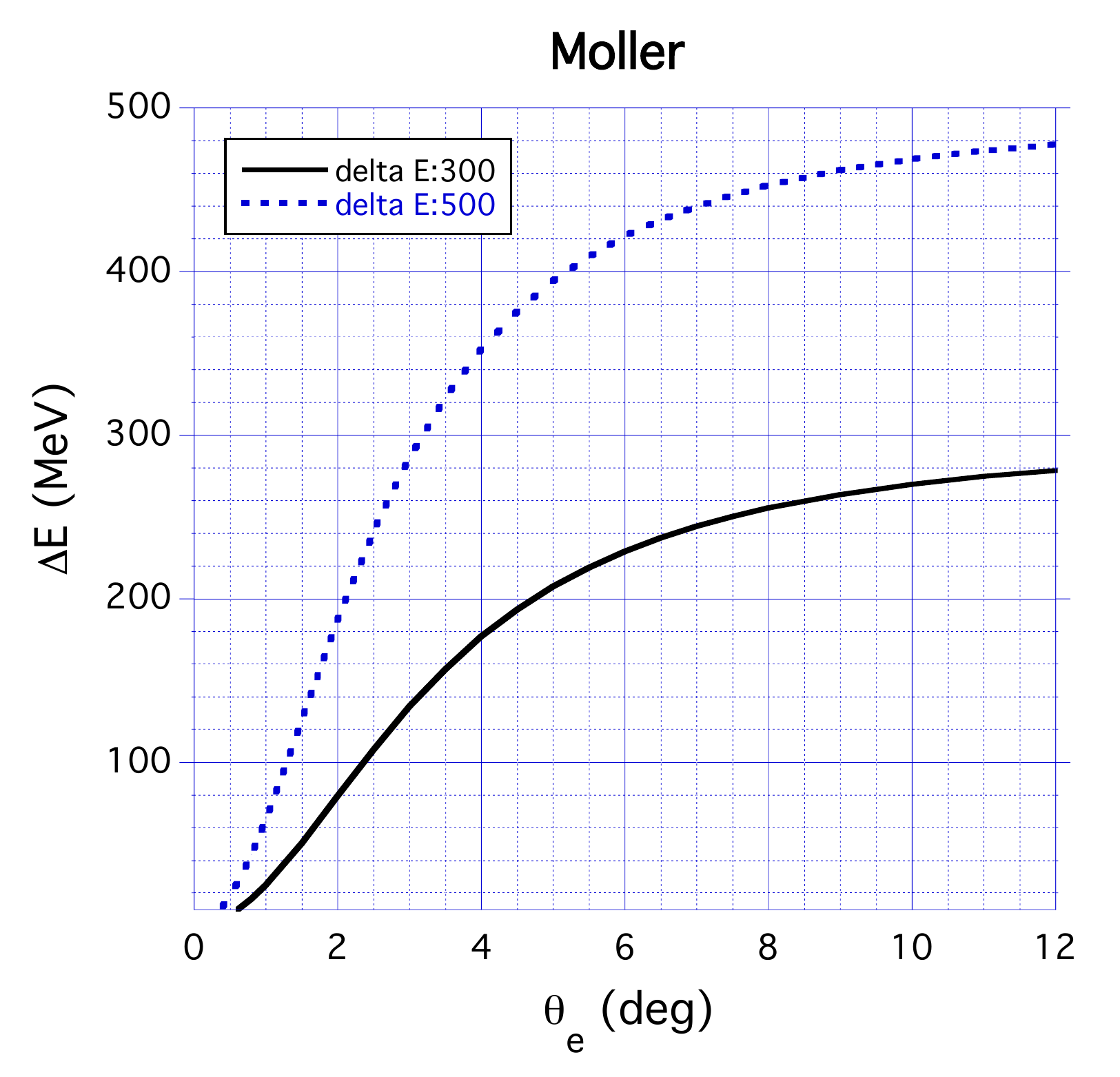}
  \includegraphics[height=.3\textheight,width = 0.45\textwidth]{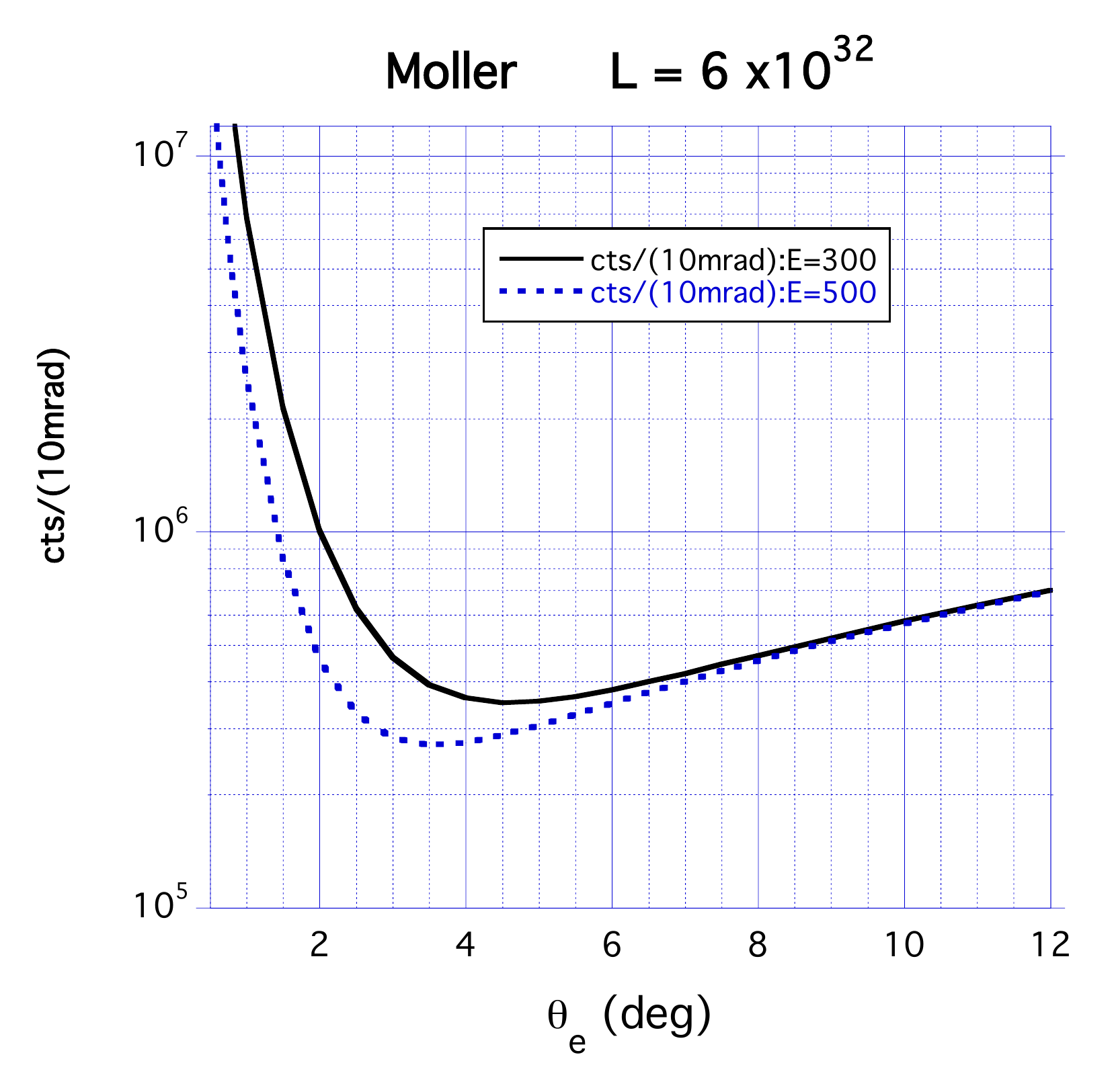}
  \caption{Moller scattering for incident energies of 300 and 500 MeV. Left Panel: energy loss $\Delta$E of scattered electrons versus $\theta_{e}$. Right panel: count rate in Hz into a 10 mrad angular ring versus $\theta_{e}$ for an electron luminosity  = 6x$10^{32} cm^{-2}sec^{-1}$ corresponding to a current of 1 mA and a target density of $10^{17} cm^{-2}$(see text for discussion). }
  \label{fig:Moller}
\end{figure}

A schematic version of small angle electron scattering is shown in Fig.\ref{fig:setup}. An electron beam is incident on a target and the scattered electrons are detected at small angles, preferably with $\phi$ symmetry to increase the counting  rate and also to enable simultaneous determination of reactions with a range of out of plane angles. 
The use of a magnetic spectrometer enables measurements of the electron energy loss. The main background rate in the electron detector is due to Moller scattering. Fig.\ref{fig:Moller} shows the energy loss $\Delta$E of the scattering electrons versus electron angle for incident energies of 300 and 500MeV and the count rate for 
the anticipated luminosity(see discussion after Eq.\ref{eq:lum}).It can be seen that for  $\Delta$E ~ 200 MeV, which is required for pion production, one needs angles $\simeq 4^{0} (2^{0})$ for incident electron energies of 300 (500) MeV in order to have the Moller electron energy loss sufficiently large to be able to eliminate this background by magnetic analysis. For smaller electron angles the rates could be handled down to $\theta_{e} \simeq 2^{0}(1^{0})$ but probably not at smaller angles. 

To a good approximation the relationship between small angle electron scattering and the photo-hadron cross section is\cite{Donnelly}
\begin{eqnarray}
d\sigma^{e,e'x}/(d\Omega_{e} | dE^{'}|  d \Omega_{x}) \simeq R_{0}/sin^{2}(\theta_{e}/2) d\sigma^{\gamma,x}/d\Omega_{x} \\  \nonumber
R_{0} =    \alpha /(8 \pi E_{\gamma}) (1+(E^{'}/E)^{2}), ~ ~ E_{\gamma} - E - E^{'}   \\ \nonumber
\Delta E_{\gamma} \int d\Omega_{e} d\sigma^{e,e'x}/(d\Omega_{e} dE^{'} d \Omega_{x}) \simeq 4~\Delta E_{\gamma}~ R_{0} ~\Delta \phi_{e}~ ln(\theta_{e}^{max}/\theta_{e}^{min} ) d\sigma^{\gamma,x}/d\Omega_{x}
 \label{eq:dsig}
\end{eqnarray}
where E(E') is the incident(final) electron energy,$\theta_{e}$ the final electron angle in the lab frame, x the produced hadron, and $E_{\gamma}$ the photon energy. For low $Q^{2}$ values the longitudinal contribution to the cross section is negligible (estimated to be less than $\sim$ 2\%) and the photons are almost real. It has been assumed that $\theta_{e} > 5 m_{e}/E(\simeq 0.5^{0}$ at E =300 MeV) but still small enough so that $sin(\theta_{e}) \simeq \theta_{e}$.To compute the count rate we have to integrate the cross section over the solid angle of the scattered electron assumed to be a ring extending from $\theta_{e}^{min}$ to $\theta_{e}^{max}$, with the result depending only mildly on their ratio. Based on Eq.\ref{eq:dsig} we can compare the luminosities of virtual tagging(vTag) with those of conventional photon tagging(tag). 
\begin{eqnarray}
L_{\gamma}^{vTag} = N_{e}^{vTag} ~R_{\gamma}^{vTag} ~N_{T}^{vTag}\simeq 8x10^{27}cm^{-2}sec{-1}\\  \nonumber
R_{\gamma}^{vTag} =   4~ R_{0}~ \Delta \phi_{e}~ ln(\theta_{e}^{max}/\theta_{e}^{min} ) \simeq 1.2x10^{-5}MeV^{-1}   \\ \nonumber
L_{\gamma}^{tag} = N_{e}^{tag}~ \epsilon_{tag} ~N_{T}^{tag}\simeq 8x10^{27}cm^{-2}sec^{-1}
 \label{eq:lum}
\end{eqnarray}
where for the virtual tagger we have assumed a beam energy of 300 MeV, a photon energy of 150 MeV, an electron current of 1 mA, integrated the outgoing electrons from 4 to 14 deg. in $\theta_{e}$ and over half of the ring in $\phi$ as shown in Fig.\ref{fig:setup}, and a target density of $10^{17}/cm^{2}$.This corresponds to an unpolarized target pressure of $\simeq$ 1 mm Hg and a target length of 4 cm. For a polarized target BLAST has achieved a target density of $\simeq 10^{14}$ protons(deuterons) per $cm^{2}$\cite{BLAST} with a target tube diameter of 1.5 cm. It is anticipated that this diameters can be reduced by an order of magnitude and this would increase the target density to the desired range(see\cite{Bernauer} for a discussion).With these parameters the photon luminosity will equal the one utilized by the A2 collaboration at Mainz[Mainz] for which $N_{e}^{tag} \simeq 5x10^{5}Hz, N_{T}^{tag} \simeq 8x10^{22} cm^{-2}, \epsilon_{Tag} \simeq 0.2$.This is essential to achieve high quality results.Virtual tagging has two main advantages over the conventional tagging technique. First and most important would be he pure, thin targets. In addition, with the magnetic elimination of the Moller electrons, the main count rate limitations are due to the maximum current and energy of the accelerator. and  the throughput of the detector system.

\begin{figure}
  \includegraphics[height=.29\textheight, width = 0.55\textwidth]
  {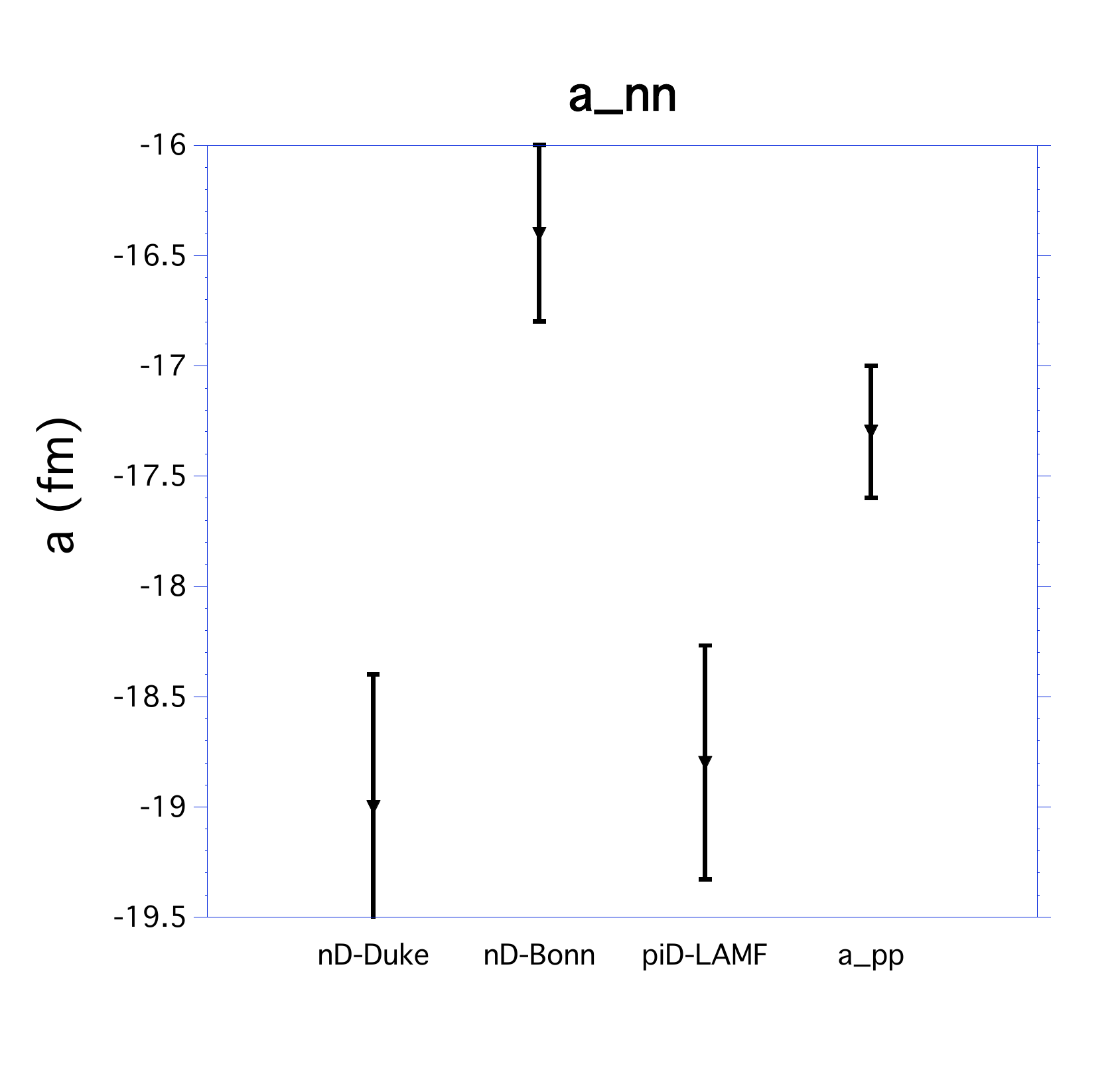}
  \includegraphics[height=.25\textheight]{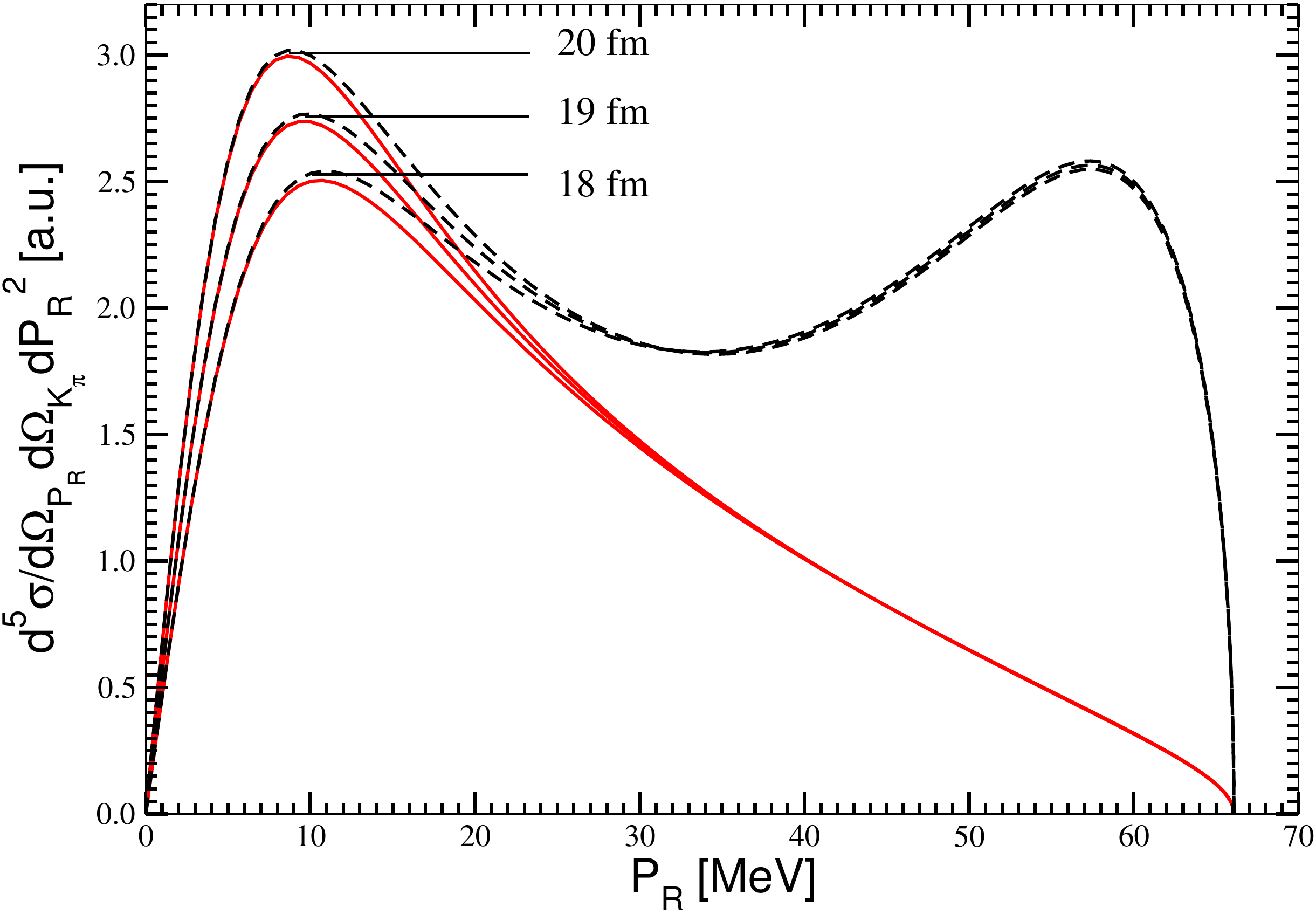}
  \caption{.Left panel: Experimental status of $a_{nn}$ and $a_{pp}$ The first two points come from the $n D \rightarrow nnp$ reaction, the third from the $\pi^{-} D \rightarrow \pi^{0}nn$ reaction(see\cite{Lensky} for references and discussion). Right panel: The theoretical cross section for the $\gamma D \rightarrow n n \pi^{+}$ reaction\cite{Lensky} versus the relative momentum of the two outgoing neutrons $p_{r}$. The cross sections are in units of $10^{-5} \mu b ~sr^{-2} MeV^{-2}$ for a CM energy of 5 MeV above threshold and a pion CM production angle $\theta_{\pi} = 135^{0}$. The solid(dashed) curves are for $\theta_{r} = 90^{0}(0^{0})$ shown for $a_{nn}$ = 18,19, and 20 fm(angles defined relative to the incident photon momentum as the z axis). See text for discussion.  }
  \label{fig:a_nn}
\end{figure}

Based on these considerations we can now discuss some of the fundamental physics possibilities in photo-pion production which are created by the development of high intensity electron beams (for the present status of the field see\cite{Hornidge,AB,Cesar}). Due to the low target density we can detect low energy charged particles from the reactions. One new possibility is the measurement of the nn s wave scattering length $a_{nn}$ from the $\gamma D \rightarrow nn \pi^{+}$ reaction just above threshold\cite{Lensky}. If charge symmetry holds then $a_{nn} = a_{pp}$. There has been a long standing experimental problem in the the determination of $a_{nn}$ which is illustrated in Fig.\ref{fig:a_nn} in that several experiments have reported significantly different results(see \cite{Lensky} for references). Thus it is not clear if charge symmetry is violated or not. It is of importance to have an independent check. To measure $a_{nn}$ from the $\gamma D \rightarrow nn \pi^{+}$ reaction one should choose kinematics where the two neutrons emerge at low relative momenta.    
 Due to chiral symmetry the $\pi$ N interaction is weak at low energies so that its interaction with the neutron pair is small. This reaction has been carefully studied using ChPT(chiral perturbation theory)\cite{Lensky} and the sensitivity to $a_{nn}$ is shown in Fig.\ref{fig:a_nn}. The curves are shown in terms of the magnitude of the relative momentum of the two outgoing neutrons  $\vec{p_{r}} =(\vec{p_{1}}- \vec{p_{2}}$)/2. The peak for small $p_{r}$ is senstive to $a_{nn}$. The peak at larger $p_{r}$ is due to  quasi-free production of the two neutrons; this peak is present when the angle $\theta_{r} = 90^{0}$ and not for $\theta_{r} = 0^{0}$(angles defined relative to the incident photon momentum).This sensitivity indicates that a kinematically complete experiment must be performed in order to obtain the full sensitivity to $a_{nn}$. It has the advantage that with one experiment the reaction calculation and the deuteron wave function can be tested by a measurement of the quasi-free peak and the magnitude of $a_{nn}$ from the low $p_{r}$ peak. Another advantage of this reaction is that it can be extended to other charge states. The well known value of $a_{np}$  can also be obtained from a measurement of the $\gamma D \rightarrow \pi^{0} np$ reaction; this will provide an excellent test of the method and the reaction calculation.  

The use of deuterium targets will also allow for the accurate measurement of pion production amplitudes from the neutron, a topic which has hardly been addressed by previous and current experiments. The value of low density targets allows for the measurement of the coherent cross section for the $\gamma D \rightarrow \pi^{0} D$ reaction by measurement of the recoil deuterons. Near threshold, to first approximation the cross section is sensitive to the square of the coherent sum of the s wave electric dipole amplitudes of the proton and neutron $| E_{0+}(\gamma p \rightarrow \pi^{0}p) + E_{0+}(\gamma n \rightarrow \pi^{0}n)|^{2}$. An accurate measurement of this cross section will allow for a determination of both the relative signs and magnitudes of these quantities for the first time and provide a stringent test of ChPT calculations. This apparatus can also be extended to make measurements at low values of the four momentum transfer $Q^{2}$.  

An important  use of deuterium (and possibly polarized $^{3}He$) targets is to obtain the neutron amplitudes in the $\gamma N \rightarrow \pi$N reactions to test chiral calculations and also, for the first time, to check isospin conservation.
\begin{eqnarray}
A(\gamma p \rightarrow \pi^{+}n) + A(\gamma n \rightarrow \pi^{-}p) = \sqrt 2[~ A(\gamma n \rightarrow \pi^{o}n) - A(\gamma p \rightarrow \pi^{0}p) ~]
 \label{eq:IS}
\end{eqnarray}
where A is any multipole amplitude ($E_{0+}, M_{1+}$,..). In principle this relationship should be modified by corrections due  to the up, down quark mass differences. This has always been assumed to be correct, but  has never been tested experimentally nor have the corrections been theoretically estimated. One  isospin breaking correction is due to $\pi^{0}, \eta, \eta^{'}$mixing which causes a 2.2\% correction to the amplitude for the $\pi^{0} \rightarrow \gamma \gamma$ decay rate\cite{AB-RMP}. There should also be additional electromagnetic corrections which show up in the neutron, proton and  charged, neutral pion mass differences. The experiments would involve extractions  the $\gamma n \rightarrow \pi^{0}n, \pi^{-}p$ cross sections and asymmetries from the $\gamma D \rightarrow \pi^{0}n, \pi^{-}p$ reactions. A clear check on the calculations of the few body reaction dynamics would be to extract the proton amplitudes and check them against the measured quantities. 

The previous examples do not require polarized targets, which if employed, could lead to other important experiments.The polarized target asymmetry T for transverse polarized targets are sensitive to the final state $\pi$N  phase shifts. Such measurements will lead to a  test of the Fermi-Watson (final state interaction) theorem. The most likely reason for a violation would be due to isospin breaking. An accurate measurement T in the near threshold region shows  the unitary cusp in the $\gamma \vec{p} \rightarrow \pi^{0}p$ reaction\cite{AB} and  will determine the s wave charge exchange scattering length  $a_{cex} (\pi^{+} n) \rightarrow \pi^{0}p)$ which can be compared to the measurement of $a_{cex}(\pi^{-}p \rightarrow \pi^{0}n)$ to test the predicted violation of isospin conservation in the $\pi$N system\cite{AB,Martin}. This can be extended to higher energies where isospin is also expected to be broken by both electromagnetic and quark mass difference dynamics. 

In conclusion, some important experiments that can be performed with a high intensity electron beam with energies $\geq$ 300 MeV have been sketched. It also may be possible to extend some of these measurements to somewhat higher photon virtualities $Q^{2}$, particularly if the beam energies are increased. This would measure the spatial extent (RMS  radius) of these transition matrix elements. To exploit this possibility will require careful design work for each experiment and close collaboration between theory and experiment. These fundamental experiments will test ChPT and its link to  QCD, and hopefully future lattice calculations. With sufficient precision isospin breaking due to the mass difference of the up and down quarks can be probed.



\begin{thebibliography}{9}
\bibitem{BLAST} D.Hasell et al., Ann.Rev.Nucl.Part.Sci. 61,409 (2011), and Nucl.Instrum.Meth. A603,247(2009)
\bibitem{POLITE} A.M. Bernstein and M.M. Pavan, Proceedings of the Second Workshop, on Electromagnetic Physics with Internal Targets and the BLAST Detector, R. Alarcon and R.Milner editors, World Scientific, Singapore(1999), A.M. Bernstein et al., Bates Internal Reports,June 2002, and March 1996. 
\bibitem{Hornidge} D. Hornidge \textit{et al.}, arXiv:1211.5495 [nucl-ex]
\bibitem{Donnelly} T. W. Donnelly and A. S. Raskin, Ann. Phys. {\bf 169}, 247(1986); 
A. S. Raskin and T. W. Donnelly, Ann. Phys. {\bf 191}, 78 (1989).
\bibitem{Bernauer} J. Bernauer and R. Milner, contribution to this workshop. 
\bibitem{AB} A.M.Bernstein, arXiv:1304.4276[hep-ex], A. M. Bernstein, M. W. Ahmed, S. Stave, Y. K. Wu, and H. Weller, Ann. Rev. Nucl. and Part. Sci., {\bf 59}, 115 (2009).
\bibitem{Cesar} M. Hilt, S. Scherer, and L. Tiator, Phys. Rev. C 87, 045204 (2013). C. Fernandez-Ramirez and A.M. Bernstein, Phys. Lett. B (2013), http://dx.doi.org/10.1016/j.physletb.2013.06.020
\bibitem{Lensky} V. Lensky et al., Eur.Phys.J. A33, 339(2007).
\bibitem{AB-RMP}A.M. Bernstein and Barry R. Holstein, Rev. Mod. Phys. 85,49(2013).
\bibitem{Martin}Martin Hoferichter, Bastian Kubis, Ulf G. Meiner, Nucl.Phys. A833,18 (2010), Phys.Lett. B678,65 (2009).
\end{thebibliography}
\end{document}